\documentclass[twocolumn,english,aps]{revtex4}
\usepackage[T1]{fontenc}
\usepackage[latin9]{inputenc}
\usepackage{textcomp}
\usepackage{amsmath}
\usepackage{graphicx}
\usepackage{amssymb}
\usepackage{esint}

\makeatletter
\@ifundefined{textcolor}{}
{%
 \definecolor{BLACK}{gray}{0}
 \definecolor{WHITE}{gray}{1}
 \definecolor{RED}{rgb}{1,0,0}
 \definecolor{GREEN}{rgb}{0,1,0}
 \definecolor{BLUE}{rgb}{0,0,1}
 \definecolor{CYAN}{cmyk}{1,0,0,0}
 \definecolor{MAGENTA}{cmyk}{0,1,0,0}
 \definecolor{YELLOW}{cmyk}{0,0,1,0}
 }

\@ifundefined{definecolor}
 {\usepackage{color}}{}
\usepackage{mciteplus}

\makeatother

\usepackage{babel}

\makeatother

\usepackage{babel}

\begin{document}

\title{Influence of Surface Waves on Plasma High Harmonic Generation}

\author{Daniel an der Brügge$^{1}$, Naveen Kumar$^{1}$, Alexander Pukhov$^{1}$,
Christian Rödel$^{2}$}

\affiliation{$^{1}$Institut für theoretische Physik I, Heinrich-Heine-Universität
Düsseldorf, D-40225 Düsseldorf}

\affiliation{$^{2}$Institut für Optik und Quantenelektronik, Friedrich-Schiller-Universität
Jena, D-07743 Jena}
\begin{abstract}
The influence of surface plasma waves (SPW) on high harmonic generation
(HHG) from the interaction of intense lasers with overdense plasma
is analyzed. It is shown, that the surface waves lead to the emission
of harmonics away from the optical axis. These off-axis harmonics
violate the parity selection rules found from 1D models. Further,
our investigations in the highly relativistic regime point towards
the existence of a new SPW generation process.
\end{abstract}
\maketitle
\global\long\def\spw{\mathrm{spw}}
\global\long\def\pic{\mathrm{PIC}}
\global\long\def\tw{2\omega}
\global\long\def\skin{\mathrm{skin}}

High harmonics generated from solid density plasma surfaces may be
about to become the tool that enables the next breakthrough in ultrafast
science \cite{tsakiris2006routeto,krausz2009attosecond}. Extremely
intense XUV attosecond pulses are emitted from plasma surfaces, exploiting
the non-linearity of relativistic physics. Because plasmas are used,
the process is not subject to the limitations of the methods for generating
attosecond pulses in gases \cite{tsakiris2006routeto}.

The generation scheme works as follows: an intense laser is incident
on a solid surface, immediately ionizing the material. The resulting
plasma reflects the light due to its overcritical density. If the
field intensity is sufficient, the reflection process becomes strongly
non-linear, yielding high harmonics together with the reflected light.
Different mechanisms have been identified to be responsible for the
high harmonic generation (HHG), some of them already apparent in the
sub-relativistic regime, while others take over in the highly relativistic
regime \cite{baeva2006theoryof,brugge2010enhanced,thaury2010highorder}.
All of these theoretical descriptions, including numerical simulations,
were formulated in simplified one-dimensional (1D) models, assuming
the incoming laser light as well as the reflected radiation to be
plane waves \cite{lichters1996shortpulse,baeva2006theoryof,brugge2010enhanced,thaury2010highorder}.
Diffraction and focusing of the extremely broadband harmonics radiation
in real 3D space can be treated separately, provided the HHG happens
{}``locally independently'' at each point in the focal plane \cite{brugge2007propagation}.

One of the first theoretical results about the surface harmonics have
been selection rules concerning the parity and the polarization of
the generated radiation \cite{lichters1996shortpulse}. In case of
normally incident light, the rule can be derived as follows: The radiation
source term is the transverse electron current in the skin layer,
$j_{y}=\rho p_{y}/\gamma$. Here, $\rho=-e(n_{e}-n_{e,0})$ is the
charge density (ions are assumed immobile), $p_{y}$ is the transverse
momentum component and $\gamma$ the relativistic $\gamma$-factor
of the electrons. As the reflected radiation contains the same Fourier
components as the source current $j_{y}$, we now examine, which Fourier
components the individual factors may contain. In the assumed 1D geometry,
due to the conservation of canonical momentum, we have $p_{y}\propto A_{y}$.
Thus, to first order, it oscillates at the laser frequency and we
may write $p_{y}=p_{\skin}e^{-i\omega t}$. Oscillations in the relativistic
$\gamma$-factor and the charge density $\rho$ on the other hand
are determined by the relativistic ponderomotive force and therefore
must have a period which corresponds to half of the laser period.
They contain only even harmonics, which we denote by $\rho=\rho_{\tw}$
and $\gamma=\gamma_{\tw}$. Consequently, the product $j_{y}=\rho p_{y}/\gamma$
contains only odd harmonics, and so does the reflected radiation.
In a similar way, selection rules can be derived for oblique incidence,
a complete overview of this was given in Tab.~1 of Ref.~\cite{lichters1996shortpulse}. 

The generation of surface plasma waves (SPWs) is a vital feature of
laser interaction with overdense plasma. A SPW can propagate long-distances
along the surface, while its amplitude falls off exponentially in
a direction perpendicular to the surface \cite{kaw1970surface,agranovich1975crystal,lee1999parametric,pitarke2007theoryof}.
The dispersion relations of the SPW reads as \cite{landau2008electrodynamics,pitarke2007theoryof}:\begin{equation}
ck_{\spw}=\sqrt{\frac{\omega_{p}^{2}-\omega_{\spw}^{2}}{\omega_{p}^{2}-2\omega_{\spw}^{2}}}\,\omega_{\spw},\label{eq:spw_dispersion}\end{equation}
where $\omega_{p}$ is the electron plasma frequency, $\omega_{\spw}$
and $k_{\spw}$ are the SPW frequency and wavenumber, respectively.
It has been known for some time that a laser impinged on the vacuum
overdense plasma surface can generate two counter-propagating SPWs,
and this process of SPW excitation was termed the two-surface waves
decay (TSWD) process \cite{yasumoto1982electromagnetic,macchi2001surface,macchi2002twosurface,kumar2007parametric}.
Physically, the ponderomotive force of the laser pulse causes density
oscillations at the surface. These density oscillations, at twice
of the laser frequency, beat with the electron oscillatory velocity,
causing the laser pulse to produce a nonlinear current density which
oscillates at one of the SPWs frequencies. For normal incidence of
the laser, both SPWs oscillate with the laser frequency itself, forming
a standing wave. The SPW becomes parametrically unstable and energy
flows into the SPW modes \cite{kumar2007parametric}. The expression
for the growth rate of the SPWs excitation (cf. Eq.~(26) in Ref.~\cite{kumar2007parametric})
reveals that this process becomes resonant if one chooses initial
plasma density to be $\omega_{p}=2\omega_{0}$, where $\omega_{p}$
is the laser frequency. The resonance occurs due to the presence of
the dielectric function $\epsilon_{2\omega_{0}}=1-\omega_{p}^{2}/4\omega_{0}^{2}$
in the denominator of the growth rate expression. This resonant excitation
of the SPWs in the laser-solid interaction can significantly change
the surface morphology. Thus, if present, the SPWs are expected to
affect HHG greatly as it occurs from the plasma surface. However,
to the best of our knowledge, the influence of SPWs on HHG has never
been investigated so far.

In this paper, we study the influence of SPWs on surface HHG on the
basis of two sample PIC simulation runs. We thereby focus on normal
incidence, as the influence of the SPWs can be pointed out most clearly
here. The first simulation is in the moderately relativistic regime,
where we have reason to believe that the analytic description of the
TSWD is approximately valid. The second is in the highly relativistic
regime, where HHG is supposed to be most efficient and we expect considerable
deviation from linear SPW theory. 

Let us begin with the first example. For this simulation, we chose
parameters similar to the ones in Ref.~\cite{macchi2001surface}.
The laser is a plane wave, normally incident on the target. Its temporal
envelope is given by a linear ramp over three laser periods, up to
a constant amplitude of $a_{0}\equiv eA/m_{e}c^{2}=1.7$. The simulation
domain is 2D, i.e. all derivatives in $z$-direction are neglected.
Further, the laser $\mathbf{E}$-field was chosen to point in $y$-direction.
The plasma density leaps from $0$ to $n_{e,0}=5n_{c}$ within one
grid step at $x=0.5\,\lambda$. Ions are fixed.

\begin{figure}
\includegraphics[width=3.375in]{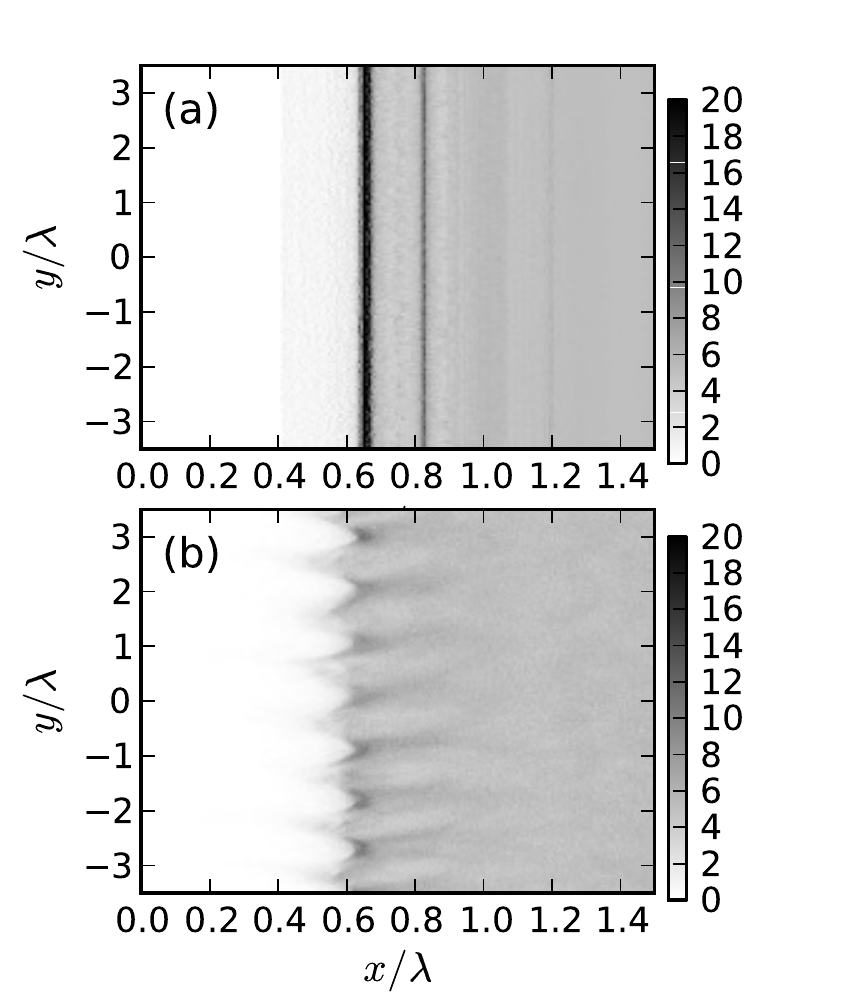}\caption{\label{fig:004_density_cartoon}Electron density profile (a) before
and (b) after the SPW emerged. Simulation parameters are: $a_{0}=1.7$,
$n_{e,0}=5n_{c}$. Frame (a) was recorded at simulation time $ct=4\lambda$,
(b) at $ct=30\lambda$.}
\end{figure}

Figure~\ref{fig:004_density_cartoon} shows the surface density profile
at two different times. Frame (a) was recorded at time $ct=4\lambda$,
just after the laser had hit the surface. Here, the electron surface,
initially located at $x=0.5\,\lambda$, has been pushed in uniformly
by about a tenth of the wavelength. Strong variations along the surface
are not observed except for some small noise, so at this point the
simulation results can still be well described by a 1D model. This
changes later in the simulation, as we can see in frame (b), recorded
26 laser cycles later. Here, a SPW has emerged, spontaneously breaking
the translational symmetry. The wavenumber, measured from the data
shown in the figure, is given by $k_{\spw}^{\pic}\approx1.1\, k_{0}$,
very close to what was expected from the linear SPW dispersion relation
Eq.~\eqref{eq:spw_dispersion}, $k_{\spw}=1.15\, k_{0}.$

\begin{figure}
\includegraphics[width=3.375in]{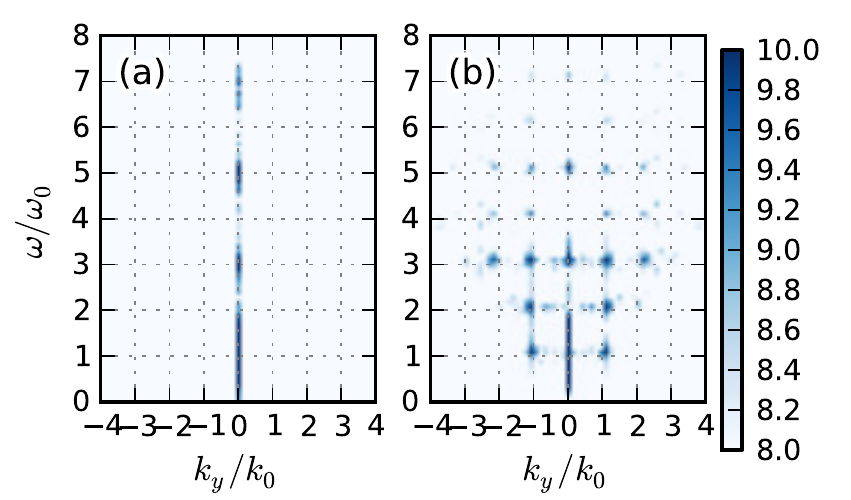}\caption{\label{fig:004_spectra}$k_{y}$-resolved harmonics spectra, recorded
at the left boundary of the simulation box. Simulation parameters
are: $a_{0}=1.7$, $n_{e,0}=5n_{c}$. Frame (a) corresponds to the
time span $ct=0\ldots10\,\lambda$, when the SPW has not yet emerged,
and (b) to $ct=25\ldots35\,\lambda$, when the SPW is fully grown.
The colour scale corresponds to the logarithm of the spectral intensity,
normalization is arbitrary, but consistent.}
\end{figure}

Corresponding spectra of the reflected light are presented in Fig.~\ref{fig:004_spectra}.
Again, frame (a) corresponds to an early state, before the SPW builds
up. It is seen, that all radiation is reflected at $k_{y}=0$, as
expected from a plane wave reflected at a smooth surface. We also
notice, that harmonics are generated at odd multiples of the laser
fundamental, in agreement with the selection rules explained above.
Also, we see that all spectral lines are distinctly broadened compared
to later times. This can in parts be understood due to the motion
of the reflecting surface, especially during the time span, when the
laser intensity is still increasing \cite{behmke2011controlling}.
Frame (b) shows the reflected radiation at the state, where the SPW
is fully developed. We see that it possesses a clear signature in
the harmonics spectrum. Now, light is not only emitted in specular
direction, but also at $k_{y}=l\, k_{\spw}$, where $l$ is an integer,
numbering the angular sideband. These wavenumbers correspond to angles
\begin{equation}
\vartheta_{m}^{l}=\arcsin\left(\frac{lck_{\spw}}{m\omega_{0}}\right)\label{eq:angles}\end{equation}
away from the optical axis, wherein $m$ is the number of the harmonic.
Thus, e.g. the 3rd harmonic is measured in specular direction, but
also at angles of $\vartheta_{3}^{1}=22\text{\textdegree}$ and $\vartheta_{3}^{2}=47\text{\textdegree}$
away from the optical axis.

Another remarkable observation is that although the specular light
still consists of purely odd-numbered harmonics, the angular sidebands
contain both even and odd harmonic numbers. This can be understood
with the following perturbative approach. As in the 1D case, the source
current is determined by three factors: the charge density $\rho$,
the transverse momentum $p_{y}$ and the relativistic factor $\gamma$.
Assuming that the SPW induces a small perturbation, we neglect its
influence on $\gamma$, writing $\gamma=\gamma_{\tw}$ as before.
To the factors $p_{y}$ and $\rho$, we add a perturbation term representing
the SPW: $\rho=\rho_{2\omega}+\delta\rho_{\spw}\cos(k_{\spw}y)\, e^{-i\omega t}$
and $p_{y}=p_{\skin}e^{-i\omega t}+\delta p_{\spw}\cos(k_{\spw}y)\, e^{-i\omega t}$.
Forming the product, we see that \begin{multline}
j_{y}=j_{y}^{0}+\cos(k_{\spw}y)\left(\frac{\delta p_{\spw}\,\rho_{\tw}}{\gamma_{\tw}}e^{-i\omega t}+\frac{\delta\rho_{\spw}\, p_{\skin}}{\gamma_{\tw}}e^{-2i\omega t}\right)\\
+\mathcal{O}(\delta^{2}),\end{multline}
wherein $j_{y}^{0}$ corresponds to the unperturbed 1D case, thus
representing odd harmonics in specular direction. The second summand
represents the first angular sideband at $k_{y}=\pm k_{\spw}$. Obviously,
both even and odd harmonics are contained here. Off-axis even harmonics
arise due to the density modulation associated with the SPW, odd ones
due to the modulation of the momentum. To assess the higher angular
sidebands, terms of the order of $\delta^{2}$ would have to be taken
into account, including the non-linearity of the SPW itself.

Figure~\ref{fig:004_spectra} contains another important detail.
We notice that when the SPW has appeared, the intensity of the higher
harmonic orders ($m=5,\,7$) is considerably depleted compared to
the 1D case. Thus, it looks as if the TSWD process competes with the
HHG process, consuming parts of its energy. This may explain why the
HHG with laser pulses in the duration range of $30\ldots100\,\mathrm{fs}$
has sometimes fallen behind expectations from 1D models and simulations.
For longer pulses, heating of the plasma suppresses the SPWs again.

For the second example, we increase both the laser amplitude and,
so as to avoid relativistic transparency, the plasma density. The
laser amplitude is raised to $a_{0}=7.2$, the plasma density to $n_{e,0}=36\, n_{c}$.
If we generously extend the resonance condition of the TSWD to allow
for excitation at harmonics instead of only the laser fundamental,
we write the dielectric term as $\epsilon_{2\omega_{m}}=1-\omega_{p}^{2}/4\omega_{m}^{2}$,
where $\omega_{m}=m\omega_{0}$. For our chosen density this term
vanishes at $m=3$, causing a TSWD resonance at the corresponding
frequency. Hence for these parameters, the dominant excitation of
the SPWs could possibly happen at the third harmonic instead of the
fundamental frequency.

\begin{figure}
\includegraphics[width=3.375in]{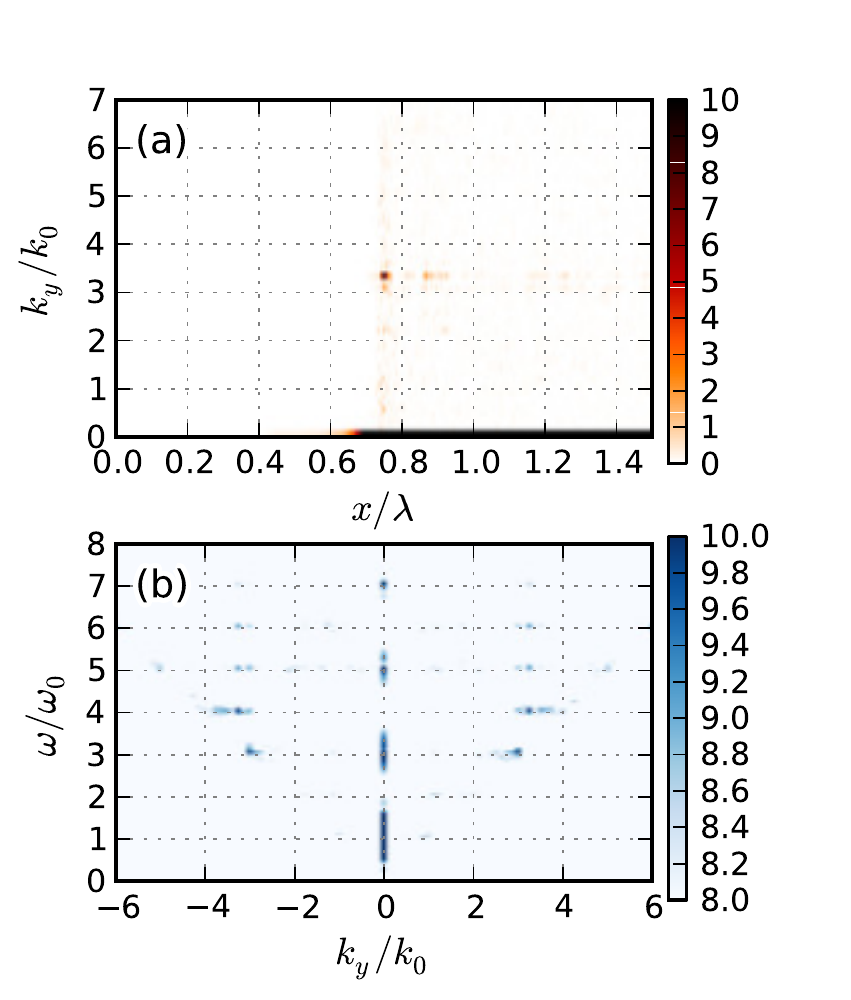}\caption{\label{fig:012_dens+spec}Results of the simulation in the highly
relativistic parameter regime: $a_{0}=7.2$, $n_{e,0}=36\, n_{c}$.
Frame (a) shows the absolute square of the Fourier transformed electron
density, recorded at $ct=18\lambda$. The colour scale is linear,
units are arbitrary. Frame (b) shows the $k_{y}$-resolved electromagnetic
spectrum, recorded at the left boundary of the simulation box, during
the time window $ct=10\ldots25\lambda$. Here, the colour scale is
logarithmic.}
\end{figure}

Simulation results are presented in Fig.~\ref{fig:012_dens+spec}.
Frame (a) shows the absolute square of the electron density, Fourier
transformed in $y$-direction. This way, SPW modes become immediately
recognizable as clearly defined spots in the image. In this example
we see that the fundamental does not efficiently excite an SPW, as
there is no spot around $k_{y}=k_{0}$. This can be expected as the
growth rate for the $\omega_{\spw}=\omega_{0}$ mode is expected to
be very small. Instead, a sharp peak is observed at $k_{y}^{\pic}\approx3.3\, k_{0}$.
This value is reasonably close to what is expected from the non-relativistic
dispersion relation for $\omega_{\spw}=3\omega_{0}$, which is $k_{y}=3.7\, k_{0}$.
We therefore conclude that here the 3rd harmonic excites the SPW instead
of the laser fundamental. To the best of our knowledge, this excitation
process is completely novel and has not yet been described in the
literature. Further simulations carried out by the authors indicate,
that this type of excitation process happens frequently in the relativistic
regime and is not limited to exact hitting of the assumed resonance.
The demonstrated sample however shows a particularly clear case.

As we see from frame (b) of Fig.~\ref{fig:012_dens+spec}, this process
also leaves its footprint in the $k_{y}$-resolved spectrum of the
reflected field. Here, no sidebands at $k_{y}=\pm k_{0}$ or $k_{y}=\pm2k_{0}$
are seen, but one at $k_{y}=\pm3k_{0}$ is clearly observed. As expected,
this sideband again contains both even and odd harmonic orders. So
e.g. the 4th harmonic order is measured at an angle of $\vartheta_{4}\approx56\text{\textdegree}$
from the optical axis, but not on the optical axis itself.

\begin{figure}
\includegraphics[width=3.375in]{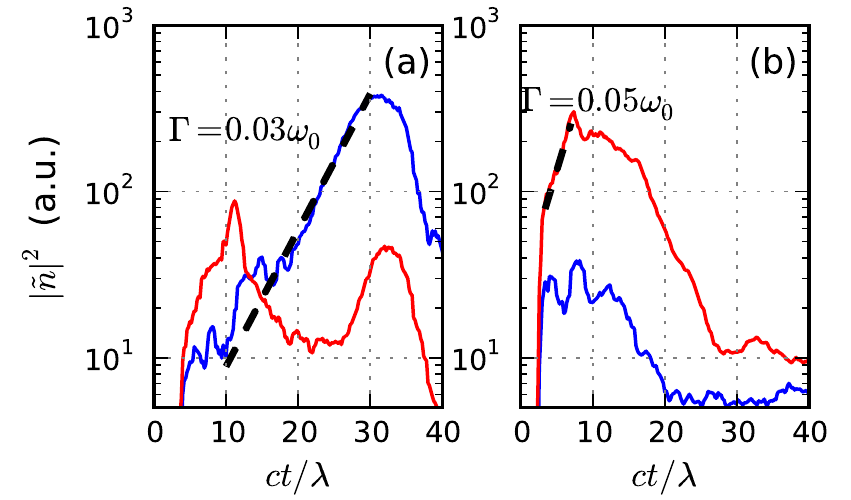}\caption{\label{fig:004_vs_012_spw_growth}Growth and damping of SPW modes,
measured as the spatial spectral intensity of density fluctuations,
defined by $|\tilde{n}|^{2}\equiv|\int n(x,y,t)\,\exp(ik_{y}y)\, dy|^{2}$.
The blue lines corresponds to modes in the range $0.9<k_{y}/k_{0}<1.3$,
for this $k$-range the SPW frequency is roughly $\omega_{0}$; the
red lines correspond to $3.0<k_{y}/k_{0}<4.0$, so that $\omega_{\spw}\approx3\omega_{0}$.
(a) shows results of the moderately relativistic simulation ($a_{0}=1.7$,
$n_{e,0}=5n_{c}$), (b) of the highly relativistic one ($a_{0}=7.2$,
$n_{e,0}=36\, n_{c}$). The black dashed lines show the exponential
fits, from which the SPW growth rate has been calculated.}
\end{figure}

Figure~\ref{fig:004_vs_012_spw_growth} shows the growth and damping
of the SPW modes in both simulations, represented by an integral over
the spatial density spectrum. Frame (a) corresponds to the moderately
relativistic simulation. It can be readily seen that after a short
period in the beginning ($ct\lesssim10\,\lambda$), where both modes
grow about equally, the SPW at the fundamental frequency (blue line)
grows strongly, whereas other modes are damped. This is expected as
we had chosen the plasma density to satisfy the plasma resonance condition
for the generation of two counter-propagating SPWs close to the fundamental
laser frequency. After the SPW has reached a certain amplitude, the
plasma is heated rapidly and the wave is damped. Results of the highly
relativistic simulation are displayed in frame (b). Here, the situation
is the other way round: Whereas the SPW at 3rd harmonic can grow quickly,
the SPW at fundamental frequency does not reach a high amplitude at
all. This can be understood, because with $n_{e,0}=36n_{c}$, the
density is detuned far from the resonance for the conventional TSWD
at the laser fundamental ($n_{e}=4n_{c}$). Consequently, the growth
rate is small and thermal damping sets in, before it can reach a significant
amplitude. On the other hand, the 3rd harmonic is exactly at resonance,
if one allows for the generous interpretation of the TSWD theory mentioned
above. Thus the SPW at this frequency can grow quickly. Note however,
that relativistic effects may have a strong bearing on the TSWD process
and the SPW dispersion relation. These qualitative discussions are
expected to be correct, however quantitative comparisons are difficult
to be drawn using the non-relativistic theory. From the PIC simulation,
we find the growth rate to be $\Gamma\approx0.05\,\omega_{0}$ within
the time range indicated in Fig.~\ref{fig:004_vs_012_spw_growth}.
Before this stage of linear growth, there is a stage of quasi immediate
growth, happening within a fraction of a laser period. After the SPW
has reached its peak, it is damped rapidly. Due to its shorter wavelength,
this SPW is even more sensitive to thermal damping compared to the
SPW at the laser fundamental frequency.

From these results, we deduce that SPW are particularly relevant for
laser pulses with the duration in a certain range. If the pulse is
too short, the SPW has no time to grow. If the pulse is too long,
electron temperature increases and SPWs are damped rapidly. The exact
duration depends on the laser and density parameters. In the cases
examined here, it was on the range of a few up to a few tens laser
periods. Note that this is a duration range, in which many of today's
high intensity laser systems work. If SPWs are present, they lead
to the production of identifiable angular sidebands in the harmonics
radiation. The angles of the sidebands can be estimated by Eq.~\eqref{eq:angles}.
As the $k_{y}$-components of the sidebands are constant, the angles
are greater for lower harmonic orders in the optical range and smaller
for high orders.

This work was supported by DFG in the framework of the TR 18 project.


\begin{thebibliography}{17}
\expandafter\ifx\csname natexlab\endcsname\relax\def\natexlab#1{#1}\fi
\expandafter\ifx\csname bibnamefont\endcsname\relax
  \def\bibnamefont#1{#1}\fi
\expandafter\ifx\csname bibfnamefont\endcsname\relax
  \def\bibfnamefont#1{#1}\fi
\expandafter\ifx\csname citenamefont\endcsname\relax
  \def\citenamefont#1{#1}\fi
\expandafter\ifx\csname url\endcsname\relax
  \def\url#1{\texttt{#1}}\fi
\expandafter\ifx\csname urlprefix\endcsname\relax\def\urlprefix{URL }\fi
\providecommand{\bibinfo}[2]{#2}
\providecommand{\eprint}[2][]{\url{#2}}

\bibitem[{\citenamefont{Tsakiris et~al.}(2006)\citenamefont{Tsakiris, Eidmann,
  {Meyer-ter-Vehn}, and Krausz}}]{tsakiris2006routeto}
\bibinfo{author}{\bibfnamefont{G.~D.} \bibnamefont{Tsakiris}},
  \bibinfo{author}{\bibfnamefont{K.}~\bibnamefont{Eidmann}},
  \bibinfo{author}{\bibfnamefont{J.}~\bibnamefont{{Meyer-ter-Vehn}}},
  \bibnamefont{and} \bibinfo{author}{\bibfnamefont{F.}~\bibnamefont{Krausz}},
  \bibinfo{journal}{New Journal of Physics} \textbf{\bibinfo{volume}{8}},
  \bibinfo{pages}{19} (\bibinfo{year}{2006}).

\bibitem[{\citenamefont{Krausz}(2009)}]{krausz2009attosecond}
\bibinfo{author}{\bibfnamefont{F.}~\bibnamefont{Krausz}},
  \bibinfo{journal}{Reviews of Modern Physics} \textbf{\bibinfo{volume}{81}},
  \bibinfo{pages}{163} (\bibinfo{year}{2009}).

\bibitem[{\citenamefont{Baeva et~al.}(2006)\citenamefont{Baeva, Gordienko, and
  Pukhov}}]{baeva2006theoryof}
\bibinfo{author}{\bibfnamefont{T.}~\bibnamefont{Baeva}},
  \bibinfo{author}{\bibfnamefont{S.}~\bibnamefont{Gordienko}},
  \bibnamefont{and} \bibinfo{author}{\bibfnamefont{A.}~\bibnamefont{Pukhov}},
  \bibinfo{journal}{Phys. Rev. E} \textbf{\bibinfo{volume}{74}},
  \bibinfo{pages}{046404} (\bibinfo{year}{2006}).

\bibitem[{\citenamefont{Br\"{u}gge and Pukhov}(2010)}]{brugge2010enhanced}
\bibinfo{author}{\bibfnamefont{D.~a.~d.} \bibnamefont{Br\"{u}gge}}
  \bibnamefont{and} \bibinfo{author}{\bibfnamefont{A.}~\bibnamefont{Pukhov}},
  \bibinfo{journal}{Physics of Plasmas} \textbf{\bibinfo{volume}{17}},
  \bibinfo{pages}{033110} (\bibinfo{year}{2010}).

\bibitem[{\citenamefont{Thaury and Qu\'{e}r\'{e}}(2010)}]{thaury2010highorder}
\bibinfo{author}{\bibfnamefont{C.}~\bibnamefont{Thaury}} \bibnamefont{and}
  \bibinfo{author}{\bibfnamefont{F.}~\bibnamefont{Qu\'{e}r\'{e}}},
  \bibinfo{journal}{Journal of Physics B: Atomic, Molecular and Optical
  Physics} \textbf{\bibinfo{volume}{43}}, \bibinfo{pages}{213001}
  (\bibinfo{year}{2010}).

\bibitem[{\citenamefont{Lichters et~al.}(1996)\citenamefont{Lichters,
  {Meyer-ter-Vehn}, and Pukhov}}]{lichters1996shortpulse}
\bibinfo{author}{\bibfnamefont{R.}~\bibnamefont{Lichters}},
  \bibinfo{author}{\bibfnamefont{J.}~\bibnamefont{{Meyer-ter-Vehn}}},
  \bibnamefont{and} \bibinfo{author}{\bibfnamefont{A.}~\bibnamefont{Pukhov}},
  \bibinfo{journal}{Phys. Plasmas} \textbf{\bibinfo{volume}{3}},
  \bibinfo{pages}{3425} (\bibinfo{year}{1996}).

\bibitem[{\citenamefont{Br\"{u}gge and Pukhov}(2007)}]{brugge2007propagation}
\bibinfo{author}{\bibfnamefont{D.~a.~d.} \bibnamefont{Br\"{u}gge}}
  \bibnamefont{and} \bibinfo{author}{\bibfnamefont{A.}~\bibnamefont{Pukhov}},
  \bibinfo{journal}{Physics of Plasmas} \textbf{\bibinfo{volume}{14}},
  \bibinfo{pages}{093104} (\bibinfo{year}{2007}).

\bibitem[{\citenamefont{Kaw}(1970)}]{kaw1970surface}
\bibinfo{author}{\bibfnamefont{P.~K.} \bibnamefont{Kaw}},
  \bibinfo{journal}{Physics of Fluids} \textbf{\bibinfo{volume}{13}},
  \bibinfo{pages}{1784} (\bibinfo{year}{1970}).

\bibitem[{\citenamefont{Agranovich}(1975)}]{agranovich1975crystal}
\bibinfo{author}{\bibfnamefont{V.~M.} \bibnamefont{Agranovich}},
  \bibinfo{journal}{Soviet Physics Uspekhi} \textbf{\bibinfo{volume}{18}},
  \bibinfo{pages}{99} (\bibinfo{year}{1975}).

\bibitem[{\citenamefont{Lee and Cho}(1999)}]{lee1999parametric}
\bibinfo{author}{\bibfnamefont{H.}~\bibnamefont{Lee}} \bibnamefont{and}
  \bibinfo{author}{\bibfnamefont{S.}~\bibnamefont{Cho}},
  \bibinfo{journal}{Physical Review E} \textbf{\bibinfo{volume}{59}},
  \bibinfo{pages}{3503} (\bibinfo{year}{1999}).

\bibitem[{\citenamefont{Pitarke et~al.}(2007)\citenamefont{Pitarke, Silkin,
  Chulkov, and Echenique}}]{pitarke2007theoryof}
\bibinfo{author}{\bibfnamefont{J.~M.} \bibnamefont{Pitarke}},
  \bibinfo{author}{\bibfnamefont{V.~M.} \bibnamefont{Silkin}},
  \bibinfo{author}{\bibfnamefont{E.~V.} \bibnamefont{Chulkov}},
  \bibnamefont{and} \bibinfo{author}{\bibfnamefont{P.~M.}
  \bibnamefont{Echenique}}, \bibinfo{journal}{Reports on Progress in Physics}
  \textbf{\bibinfo{volume}{70}}, \bibinfo{pages}{1} (\bibinfo{year}{2007}).

\bibitem[{\citenamefont{Landau et~al.}(2008)\citenamefont{Landau, Lifshitz, and
  Pitaevskii}}]{landau2008electrodynamics}
\bibinfo{author}{\bibfnamefont{L.}~\bibnamefont{Landau}},
  \bibinfo{author}{\bibfnamefont{E.~M.} \bibnamefont{Lifshitz}},
  \bibnamefont{and} \bibinfo{author}{\bibfnamefont{L.~P.}
  \bibnamefont{Pitaevskii}}, \emph{\bibinfo{title}{Electrodynamics of
  continuous media}} (\bibinfo{publisher}{{Elsevier/Butterworth} Heinemann},
  \bibinfo{address}{Amsterdam [etc.]}, \bibinfo{year}{2008}),
  \bibinfo{edition}{2nd} ed., ISBN \bibinfo{isbn}{9780750626347}.

\bibitem[{\citenamefont{Yasumoto and
  Noguchi}(1982)}]{yasumoto1982electromagnetic}
\bibinfo{author}{\bibfnamefont{K.}~\bibnamefont{Yasumoto}} \bibnamefont{and}
  \bibinfo{author}{\bibfnamefont{T.}~\bibnamefont{Noguchi}},
  \bibinfo{journal}{Journal of Applied Physics} \textbf{\bibinfo{volume}{53}},
  \bibinfo{pages}{208} (\bibinfo{year}{1982}).

\bibitem[{\citenamefont{Macchi et~al.}(2001)\citenamefont{Macchi, Cornolti,
  Pegoraro, Liseikina, Ruhl, and Vshivkov}}]{macchi2001surface}
\bibinfo{author}{\bibfnamefont{A.}~\bibnamefont{Macchi}},
  \bibinfo{author}{\bibfnamefont{F.}~\bibnamefont{Cornolti}},
  \bibinfo{author}{\bibfnamefont{F.}~\bibnamefont{Pegoraro}},
  \bibinfo{author}{\bibfnamefont{T.}~\bibnamefont{Liseikina}},
  \bibinfo{author}{\bibfnamefont{H.}~\bibnamefont{Ruhl}}, \bibnamefont{and}
  \bibinfo{author}{\bibfnamefont{V.}~\bibnamefont{Vshivkov}},
  \bibinfo{journal}{Physical Review Letters} \textbf{\bibinfo{volume}{87}}
  (\bibinfo{year}{2001}).

\bibitem[{\citenamefont{Macchi et~al.}(2002)\citenamefont{Macchi, Cornolti, and
  Pegoraro}}]{macchi2002twosurface}
\bibinfo{author}{\bibfnamefont{A.}~\bibnamefont{Macchi}},
  \bibinfo{author}{\bibfnamefont{F.}~\bibnamefont{Cornolti}}, \bibnamefont{and}
  \bibinfo{author}{\bibfnamefont{F.}~\bibnamefont{Pegoraro}},
  \bibinfo{journal}{Physics of Plasmas} \textbf{\bibinfo{volume}{9}},
  \bibinfo{pages}{1704} (\bibinfo{year}{2002}).

\bibitem[{\citenamefont{Kumar and Tripathi}(2007)}]{kumar2007parametric}
\bibinfo{author}{\bibfnamefont{N.}~\bibnamefont{Kumar}} \bibnamefont{and}
  \bibinfo{author}{\bibfnamefont{V.~K.} \bibnamefont{Tripathi}},
  \bibinfo{journal}{Physics of Plasmas} \textbf{\bibinfo{volume}{14}},
  \bibinfo{pages}{103108} (\bibinfo{year}{2007}).

\bibitem[{\citenamefont{Behmke et~al.}(2011)\citenamefont{Behmke, an~der
  Br\"{u}gge, R\"{o}del, Cerchez, Hemmers, Heyer, J\"{a}ckel, K\"{u}bel,
  Paulus, Pretzler et~al.}}]{behmke2011controlling}
\bibinfo{author}{\bibfnamefont{M.}~\bibnamefont{Behmke}},
  \bibinfo{author}{\bibfnamefont{D.}~\bibnamefont{an~der Br\"{u}gge}},
  \bibinfo{author}{\bibfnamefont{C.}~\bibnamefont{R\"{o}del}},
  \bibinfo{author}{\bibfnamefont{M.}~\bibnamefont{Cerchez}},
  \bibinfo{author}{\bibfnamefont{D.}~\bibnamefont{Hemmers}},
  \bibinfo{author}{\bibfnamefont{M.}~\bibnamefont{Heyer}},
  \bibinfo{author}{\bibfnamefont{O.}~\bibnamefont{J\"{a}ckel}},
  \bibinfo{author}{\bibfnamefont{M.}~\bibnamefont{K\"{u}bel}},
  \bibinfo{author}{\bibfnamefont{G.}~\bibnamefont{Paulus}},
  \bibinfo{author}{\bibfnamefont{G.}~\bibnamefont{Pretzler}},
  \bibnamefont{et~al.}, \bibinfo{journal}{Physical Review Letters}
  \textbf{\bibinfo{volume}{106}} (\bibinfo{year}{2011}).

\end{thebibliography}
\end{document}